\documentclass[aps,prd,twocolumn,superscriptaddress,
]{revtex4-1}
\usepackage{graphicx}
\usepackage{subcaption}
\usepackage{amsfonts}
\usepackage{amsmath}
\usepackage{comment}
\usepackage{hyperref}
\usepackage[all]{hypcap}
\usepackage[dvipsnames]{xcolor}
\hypersetup{
colorlinks=true,
linkcolor=blue,
citecolor=blue,
urlcolor=blue
}

\usepackage[
        retain-explicit-plus=true,
        retain-unity-mantissa=false,
        range-phrase=--,
        range-units=single
    ]{siunitx}
\usepackage{caption}
\def\i{\mathrm{i}}
\def\e{\mathrm{e}}

\newcommand{\psub}[1]{{\bf #1:}\\}
\renewcommand{\psub}[1]{}

\newcommand{\VarWithSub}[2]{#1_{\mathrm{#2}}}

\def\alphaFC{\VarWithSub{\alpha}{fc}}

\def\Ntot{\VarWithSub{N}{tot}}
\def\Tin{\VarWithSub{T}{in}}

\def\lengthFC{\VarWithSub{L}{fc}}
\def\Larm{\VarWithSub{L}{arm}}
\def\loss{\Lambda}

\def\omegaCarrier{\VarWithSub{\omega}{0}}
\def\Parm{\VarWithSub{P}{arm}}
\def\Larm{\VarWithSub{L}{arm}}
\def\gammaIfo{\VarWithSub{\gamma}{ifo}}

\def\K{\mathcal{K}}

\def\omegaFC{\VarWithSub{\omega}{fc}}
\def\OmegaSQL{\VarWithSub{\Omega}{SQL}}

\def\rFC{\VarWithSub{r}{fc}}

\def\Csq{\VarWithSub{C}{sqz}}
\def\Canti{\VarWithSub{C}{anti}}
\def\Closs{\VarWithSub{C}{loss}}

\newcommand{\cavityLength}{\SI{16}{\meter}}
\newcommand{\aplusLength}{\SI{300}{\meter}}

\newcommand{\armPower}{\SI{750}{\kilo\watt}}
\newcommand{\OFarmPower}{\SI{400}{\kilo\watt}}

\newcommand{\ifoBandwidth}{\SI{427}{\hertz}}
\newcommand{\ifoBandwidthSRMchange}{\SI{751}{\hertz}}

\newcommand{\sqlFreq}{\SI{65}{\hertz}}
\newcommand{\sqlFreqSRMchange}{\SI{49}{\hertz}}

\newcommand{\mitLoss}{\SI{19}{ppm}}

\newcommand{\aplusLoss}{\SI{60}{ppm}}

\newcommand{\aplusT}{\SI{1000}{ppm}}

\newcommand{\mitDetuningNoise}{\SI{12}{\hertz}}

\newcommand{\srmLowPower}{\SI{35}{\percent}}
\newcommand{\srmDesign}{\SI{20}{\percent}}

\begin{document}

\title{Optimal detuning for quantum filter cavities}

\author{Chris Whittle}
\email[]{chris.whittle@ligo.org}
\affiliation{LIGO Laboratory, Massachusetts Institute of Technology, Cambridge, Massachusetts 02139, USA}
\author{Kentaro Komori}
\email[]{kentarok@mit.edu}
\affiliation{LIGO Laboratory, Massachusetts Institute of Technology, Cambridge, Massachusetts 02139, USA}
\author{Dhruva Ganapathy}
\affiliation{LIGO Laboratory, Massachusetts Institute of Technology, Cambridge, Massachusetts 02139, USA}
\author{Lee McCuller}
\affiliation{LIGO Laboratory, Massachusetts Institute of Technology, Cambridge, Massachusetts 02139, USA}
\author{Lisa Barsotti}
\affiliation{LIGO Laboratory, Massachusetts Institute of Technology, Cambridge, Massachusetts 02139, USA}
\author{Nergis Mavalvala}
\affiliation{LIGO Laboratory, Massachusetts Institute of Technology, Cambridge, Massachusetts 02139, USA}
\author{Matthew Evans}
\affiliation{LIGO Laboratory, Massachusetts Institute of Technology, Cambridge, Massachusetts 02139, USA}

\date{\today}

\begin{abstract}
Vacuum quantum fluctuations impose a fundamental limit on the sensitivity of gravitational-wave interferometers, which rank among the most sensitive precision measurement devices ever built. The injection of conventional squeezed vacuum reduces quantum noise in one quadrature at the expense of increasing noise in the other. While this approach improved the sensitivity of the Advanced LIGO and Advanced Virgo interferometers during their third observing run (O3), future improvements in arm power and squeezing levels will bring radiation pressure noise to the forefront. Installation of a filter cavity for frequency-dependent squeezing provides broadband reduction of quantum noise through the mitigation of this radiation pressure noise, and it is the baseline approach planned for all of the future gravitational-wave detectors currently conceived. The design and operation of a filter cavity requires careful consideration of interferometer optomechanics as well as squeezing degradation processes. In this paper, we perform an in-depth analysis to determine the optimal operating point of a filter cavity. We use our model alongside numerical tools to study the implications for filter cavities to be installed in the upcoming ``A+" upgrade of the Advanced LIGO detectors.
\end{abstract}

\maketitle

\section{Introduction}
\psub{Current issue of frequency independent squeezing}
Gravitational-wave detectors represent the state-of-the-art in precision metrology. Along with other high-precision optomechanical experiments~\cite{Teufel2011, Chan2011, Wollman2015, Pirkkalainen2015, Peterson2016, Cripe2019}, they comprise a class of extremely sensitive, quantum-noise-limited systems~\cite{Aasi2015,Acernese2015,Aso2013,Grote2010}. Reduction of fundamental quantum noise with the injection of squeezed light~\cite{Caves1981} has been recently demonstrated in gravitational-wave detectors~\cite{Abadie2011, Aasi2013, Grote2013}. This technique has seen improvements in sensitivity of up to $\sim$\SI{3}{\decibel} in Advanced LIGO~\cite{Tse2019} and Virgo~\cite{Acernese2019} and \SI{6}{\decibel} in GEO600~\cite{lough2020}, enabling quantum-enhanced detection of gravitational-wave events. Unfortunately, Heisenberg's uncertainty principle demands a corresponding increase in noise in the quadrature orthogonal to the squeezing---quantum radiation pressure noise. 
Quantum radiation pressure noise already limits the useful squeezing level in current gravitational-wave detectors~\cite{Yu2020, Acernese2020}. Further increases in laser power and squeezing level---such as those planned to be implemented in the upcoming Advanced LIGO upgrade, ``A+"~\cite{barsotti2018a+}---will make its impact even larger.


\psub{Frequency-dependent squeezing}
Frequency-dependent squeezing can overcome these limitations~\cite{Kimble2001, Evans2013, Miller2015} by suppressing both quantum radiation pressure and shot noises, thereby achieving a broadband reduction of quantum noise to surpass the standard quantum limit (SQL)~\cite{braginsky1996}. A frequency-dependent rotation of the squeezing angle may be realized using the frequency response of an overcoupled cavity on reflection, called a filter cavity. The filter cavity resonance is detuned with an offset relative to the carrier frequency of the squeezed vacuum state, thus impressing a differential phase shift to the upper and lower sidebands of the squeezed field. Since the squeezed quadrature angle at a given sideband frequency is determined by the relative phase between the upper and lower sidebands, the filter cavity rotates the squeezed state for sideband frequencies that lie within the cavity linewidth. 

\psub{Past work and A+}
Table-top filter cavities have demonstrated frequency-dependent squeezing in the MHz~\cite{Chelkowski2005} and kHz regions~\cite{Oelker2016}, with longer protptype filter cavities recently achieving frequency-dependent squeezing at \SI{100}{Hz} and below~\cite{McCuller2020, Zhao2020}. 
Major upgrades to existing gravitational-wave detectors, like A+ and Advanced Virgo+, will include a filter cavity. Concepts for future gravitational-wave detectors~\cite{CEwhite,CE2015,ETdesign,ET2020} also rely on filter cavities to achieve a broadband reduction of quantum noise.
Appropriately choosing the filter cavity detuning is crucial to achieve the correct rotation of the squeezed vacuum state to match the interferometer response~\cite{Kimble2001} and maximize the benefit from frequency-dependent squeezing. 




\psub{What is new in this paper}
Here we study the optimal detuning of filter cavities for gravitational-wave detection.
Previous works have considered a variety of schemes for frequency-dependent squeezing~\cite{Kimble2001,Corbitt2004,khalili2007,khalili2008,khalili2009} and compared them with numerical optimization of filter cavity parameters as a function of length~\cite{khalili2010}.
Our work focuses on the single input filter cavity scheme being considered for current and future gravitational-wave detectors.
We present new, concise forms for the interferometer quantum noise expressions. We also run in-depth numerical optimization for the \aplusLength{} A+ filter cavity currently being installed, including both optical loss and detuning fluctuations, the main degradation mechanisms that limit frequency-dependent squeezing~\cite{Kwee2014}.
This work allows us to answer many important questions in the quest to deliver frequency-dependent squeezing to current-era gravitational-wave detectors: how to choose the filter cavity detuning for given round-trip losses, how to adapt the filter cavity detuning to changing interferometer arm powers, and how to optimize the filter cavity parameters to maximize the benefit from frequency-dependent squeezing.
While these questions are easily answerable given ideal conditions and perfect rotation matching, incorporating realistic degradations with frequency-dependent effects muddies the picture.
In this paper, we aim to bridge the gap between previous theoretical treatments of frequency-dependent squeezing and how we optimize a realistic interferometer for gravitational-wave detection.


\psub{Summary of the paper contents}
We begin by giving a general description of the optimal filter cavity detuning for frequency-dependent squeezing. We first derive concise analytical formulae for the quantum noise spectrum with and without optimal phase matching between the filter cavity and interferometer, incorporating the effects of round-trip loss and detuning fluctuations in the filter cavity. Next, we allow the input transmissivity and round-trip loss of the filter cavity to vary freely and calculate the corresponding optimal detunings from an integrated quantum noise spectrum. Finally, we apply our findings to A+. 

\section{Model}
We start by calculating the optimal input mirror transmissivity and optimal detuning for a low-loss filter cavity. Here, low loss implies a round-trip loss $\loss$ much smaller than the input mirror transmissivity $\Tin$, i.e. $\loss \ll \Tin$.

\begin{figure*}
\includegraphics[width=\hsize]{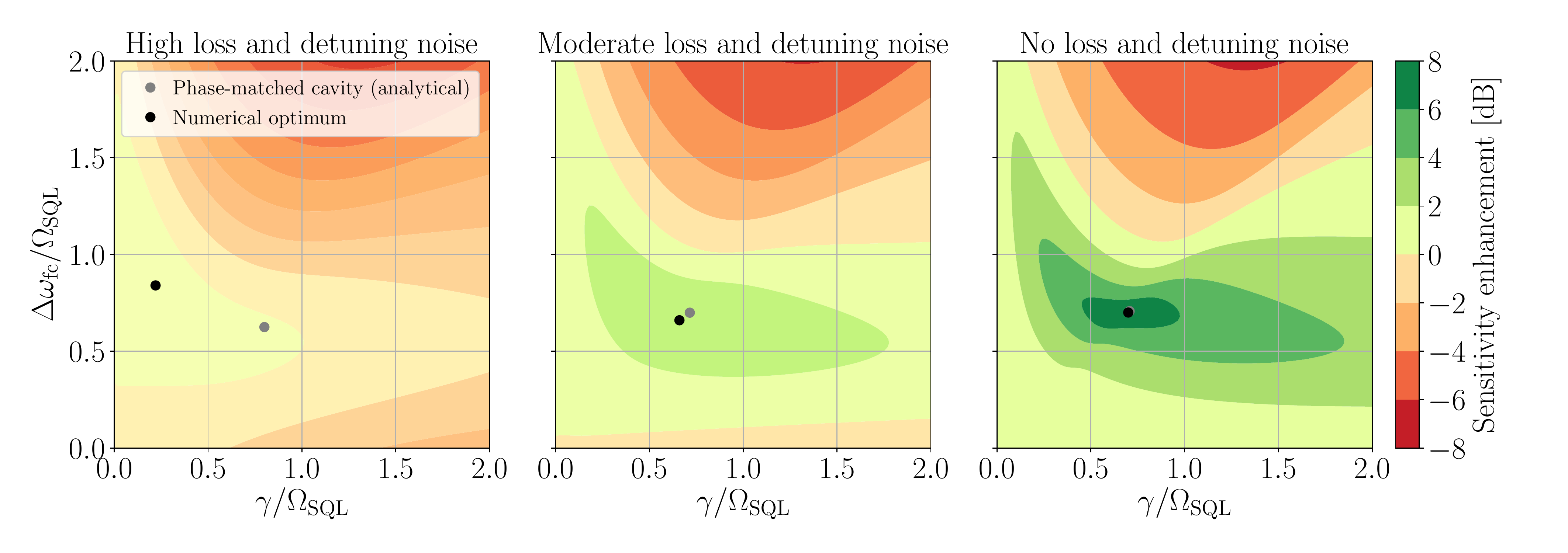}
\caption{
Sensitivity enhancement for the detection of binary inspirals relative to using frequency-independent squeezing for various cavity detunings and input transmissivities. From left to right, the three plots have round-trip losses and detuning fluctuations of $(\ell, \xi)=(0.5,0.2)$, $(0.15,0.1)$, and $(0,0)$. The enhancement factors are calculated by integrating the quantum noise spectrum over frequency, weighted by the gravitational-wave spectrum (Eq.~\ref{eq:integration}). The injected squeezing level is \SI{10}{\decibel} ($\e^{-2\sigma}=0.1$).
The gray point in each plot shows the input transmissivity and detuning calculated from the phase-matching condition (Eq. \ref{eq:optgammadet}), which assumes low loss, for the given parameters.
The black points mark the numerical maxima.
}   
\label{figure1}
\end{figure*}

\psub{Phase matching}
For a given sideband frequency $\Omega$, the optical field reflectivity of the filter cavity can be written as
\begin{equation}
\label{eq:rfcgen}
\rFC(\Omega) = 1 - \frac{2 \gamma}{\gamma + \lambda + \i(\Omega - \Delta \omegaFC)},
\end{equation}
where
\begin{equation}
\label{eq:gammalambda}
\gamma = \frac{c\Tin}{4\lengthFC},\ \lambda = \frac{c\loss}{4\lengthFC}
\end{equation}
are the \emph{coupler-limited bandwidth} and \emph{loss-limited bandwidth} respectively.
The loss-limited bandwidth is related to the frequently-used loss-per-length, differing only by a factor of $c/4$.
$\Delta \omegaFC$ is the cavity detuning with respect to the carrier frequency. The phase of a reflected sideband is given by
\begin{equation}
\alphaFC(\Omega) = \arctan{\left( \frac{2\gamma(\Omega - \Delta \omegaFC)}{-\gamma^2 + \lambda^2 + (\Omega - \Delta \omegaFC)^2} \right)}.
\end{equation}
The quadrature rotation angle of the input light after reflection from the cavity is calculated as
\begin{align}
\alpha_{\mathrm{p}} &= \frac{\alphaFC(+\Omega) + \alphaFC(-\Omega)}{2} \nonumber \\
&= \frac{1}{2} \arctan{\left( \frac{A}{B + 2(\gamma^2 + \lambda^2)\Omega^2} \right)},
\end{align}
using
\begin{gather}
A = 4\gamma \Delta \omegaFC(\gamma^2 - \lambda^2 + \Omega^2 - \Delta \omegaFC^2), \\
B = (-\gamma^2 + \lambda^2)^2 + 2(-3\gamma^2 + \lambda^2)\Delta \omegaFC^2 + (\Omega^2 - \Delta \omegaFC^2)^2.
\end{gather}
From our low-loss assumption, the coupler-limited bandwidth is much larger than the loss-limited one ($\gamma \gg \lambda$). This permits rewriting $\gamma^2 + \lambda^2 \simeq \gamma^2 - \lambda^2$ and subsequently simplifies the expression for the rotation significantly:
\begin{align}
\label{eq:sidebandrotation}
\alpha_{\mathrm{p}} &\simeq \frac{1}{2} \arctan{\left( \frac{A}{B + 2(\gamma^2 - \lambda^2)\Omega^2} \right)} \nonumber \\
&= \arctan{\left( \frac{2\gamma \Delta \omegaFC}{\gamma^2 - \lambda^2 + \Omega^2 - \Delta \omegaFC^2} \right)}.
\end{align}
The objective of the filter cavity is to apply an appropriate rotation to the squeezed state such that the ponderomotive action of the interferometer results in reduced noise rather than enhanced radiation pressure noise~\cite{Kimble2001}. The ponderomotive effect of the interferometer is characterized by the interaction strength 
\begin{align}
    \K = & \left(\frac{\OmegaSQL}{\Omega}\right)^2 \frac{\gammaIfo^2}{\Omega^2 + \gammaIfo^2}\\
    \simeq & \frac{\OmegaSQL^2}{\Omega^2},
\end{align}
where the standard quantum limit (SQL) frequency $\OmegaSQL$ is the scale factor at which $\K \simeq 1$ and is characteristic of the interferometer configuration (see Eq.~\ref{eq:OmegaSQL}).
The approximation for $\K$ holds when the interferometer bandwidth $\gammaIfo \gg \OmegaSQL$, in which case a single filter cavity is sufficient for achieving optimal rotation~\cite{khalili2010}.
The desired rotation to counteract this is then given by $\alpha_{\mathrm{p}}=\arctan\left(\K \right)$.
Setting the cavity rotation equal to that required to cancel the ponderomotive rotation gives
\begin{equation}
\arctan{\left( \frac{2\gamma \Delta \omegaFC}{\gamma^2 - \lambda^2 + \Omega^2 - \Delta \omegaFC^2} \right)} = \arctan{\left( \frac{\OmegaSQL^2}{\Omega^2} \right)}.
\end{equation}
Solving this equation finds the filter cavity parameters required for phase matching at all frequencies:
\begin{equation}
\begin{cases}
2\gamma \Delta \omegaFC = \OmegaSQL^2 \\
\gamma^2 - \lambda^2 - \Delta \omegaFC^2 = 0.
\end{cases}
\end{equation}
Thus, the optimal input transmissivity and detuning for matching a low-loss filter cavity to an interferometer with a known $\OmegaSQL$ can be written in terms of the filter cavity loss-limited bandwidth $\lambda$ as
\begin{equation}
\label{eq:optgammadet}
\begin{cases}
\gamma = \sqrt{\cfrac{\lambda^2 + \sqrt{\lambda^4 + \OmegaSQL^4}}{2}} \\
\Delta \omegaFC = \sqrt{\cfrac{-\lambda^2 + \sqrt{\lambda^4 + \OmegaSQL^4}}{2}}.
\end{cases}
\end{equation}
Eq.~\ref{eq:optgammadet} explicitly states the optimal phase-matching conditions of a filter cavity in the low-loss limit.
However, the generality of these equations remain in question.
Further, given some fixed $\gamma$ that does not necessarily follow Eq.~\ref{eq:optgammadet}, how should $\Delta \omegaFC$ be refined?
We aim to address these concerns in the rest of this paper.

\psub{Squeezing degradation with loss and detuning noise}
Next, we go beyond the low-loss limit and optimal phase-matching conditions to derive the sensitivity enhancement from squeezing by a factor $\e^{-\sigma}$ in a more general parameter space. We break this calculation down into the contributions of the squeezing, anti-squeezing, and unsqueezed vacuum terms to the total quantum noise.

Defining two new parameters as
\begin{gather}
\mu = \frac{\rFC(+\Omega) + \rFC^*(-\Omega)}{2}, \\
\nu = -\i \frac{\rFC(+\Omega) - \rFC^*(-\Omega)}{2},
\end{gather}
the power from quantum noise measured at the detection port of a gravitational-wave interferometer can be written as
\begin{equation}
\label{eq:Ntots}
\Ntot = \Csq \e^{-2\sigma} + \Canti \e^{2\sigma} + \Closs,
\end{equation}
where 
\begin{gather}
\label{eq:sqcoeff1}
\Csq = |\K \nu + \mu|^2, \\
\label{eq:sqcoeff2}
\Canti = |-\K \mu + \nu|^2, \\
\label{eq:sqcoeff3}
\Closs = \left( 1 - |\mu|^2 - |\nu|^2 \right) \left( \K^2 + 1\right).
\end{gather}
These equations can be used to estimate the enhancement for any input transmissivity, loss, and detuning.
We henceforth refer to these results as our analytical model.

We now numerically compute the overall sensitivity enhancement resulting from frequency-dependent squeezing as a function of these three parameters: $\gamma/\OmegaSQL$, $\ell \equiv \lambda/\OmegaSQL$ and $\Delta \omegaFC/\OmegaSQL$. The enhancement factor is defined using a frequency-weighted integral of $N_{\mathrm{tot}}$ from Eq.~\ref{eq:Ntots}, normalized by the same frequency-weighted integration of quantum noise with frequency-independent squeezing. It can be written as
\begin{equation}
    \label{eq:integration}
    I = \cfrac{\int_0^{\infty} d\Omega\; \left[ \Omega^{-7/3} N_{\mathrm{tot}}^{-1}\, \K \Omega^2 \right]} {\int_0^{\infty} d\Omega\; \left[ \Omega^{-7/3} (\K^2 \e^{2\sigma} + \e^{-2\sigma})^{-1}\, \K \Omega^2 \right]}.
\end{equation}
An integration weighted by $\Omega^{-7/3}$ can be used as a proxy for the enhancement of the signal-to-noise (SNR) ratio for gravitational waves detected from a binary inspiral.
The $-7/3$ frequency exponent comes directly from the power spectrum of a binary inspiral~\cite{maggiore2007}.
In these calculations, we now also incorporate detuning fluctuations as a squeezed state degradation mechanism. These fluctuations arise from residual cavity length noise and create a form of frequency-dependent phase noise. We give this noise in a dimensionless form $\xi = \delta \omegaFC / \OmegaSQL$ by normalizing the detuning fluctuation $\delta \omegaFC$ by the SQL frequency. See also Appendix~\ref{app:formula} for an extension of our analytical model that incorporates these effects.

\psub{Detuning v.s. input transmissivity}
Fig.~\ref{figure1} shows the enhancement factor with varying $\Delta \omegaFC$ and $\gamma$, while fixing the loss $\ell$ and detuning fluctuation $\xi$.
An optical loss as large as $\ell = 0.5$ limits the sensitivity enhancement to below \SI{1}{\decibel}, and should remain smaller than 0.15 in order to achieve around \SI{3}{\decibel}. When $\ell \lesssim 0.15$ is enforced, the optimal input transmissivity and detuning given by Eq.~\ref{eq:optgammadet}---for which the low-loss and low-noise limit is assumed---is very close to the numerical maximum.

\psub{Detuning v.s. loss}
\begin{figure*}
\centering
\begin{minipage}[t]{0.48\hsize}
\vspace{0pt}
\centering
\includegraphics[width=\hsize]{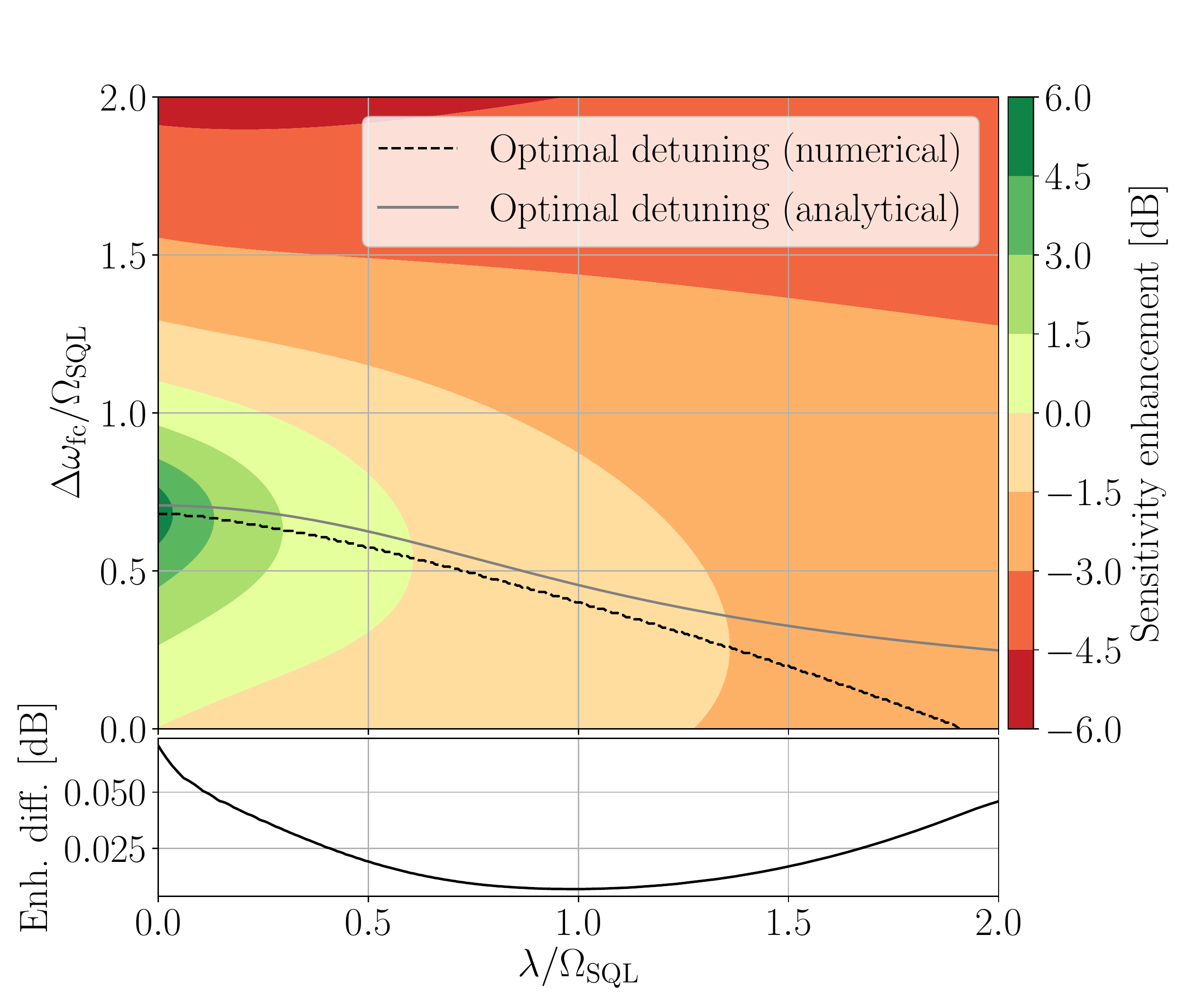}
\caption{
Sensitivity enhancement factor for binary inspirals (Eq.~\ref{eq:integration}) using filter cavities with various detunings and losses at a given input transmissivity $\gamma/\OmegaSQL=1/\sqrt{2}$ and detuning fluctuation $\xi=0.1$. The black and gray lines represent the best detuning at a given optical loss from numerical optimization and Eq.~\ref{eq:optgammadet} respectively.
The lower panel shows the ratio between the numerically-optimized and analytically-optimized sensitivity enhancements, calculated along each of the curves in the top panel.
The optimal detuning decreases with larger losses. We also find that the detuning derived from the analytical equation gives almost the same factor of improvement as the numerical maximum.
}   
\label{figure2}
\end{minipage}\hfill
\begin{minipage}[t]{0.48\hsize}
\vspace{0pt}
\centering
\includegraphics[width=\hsize]{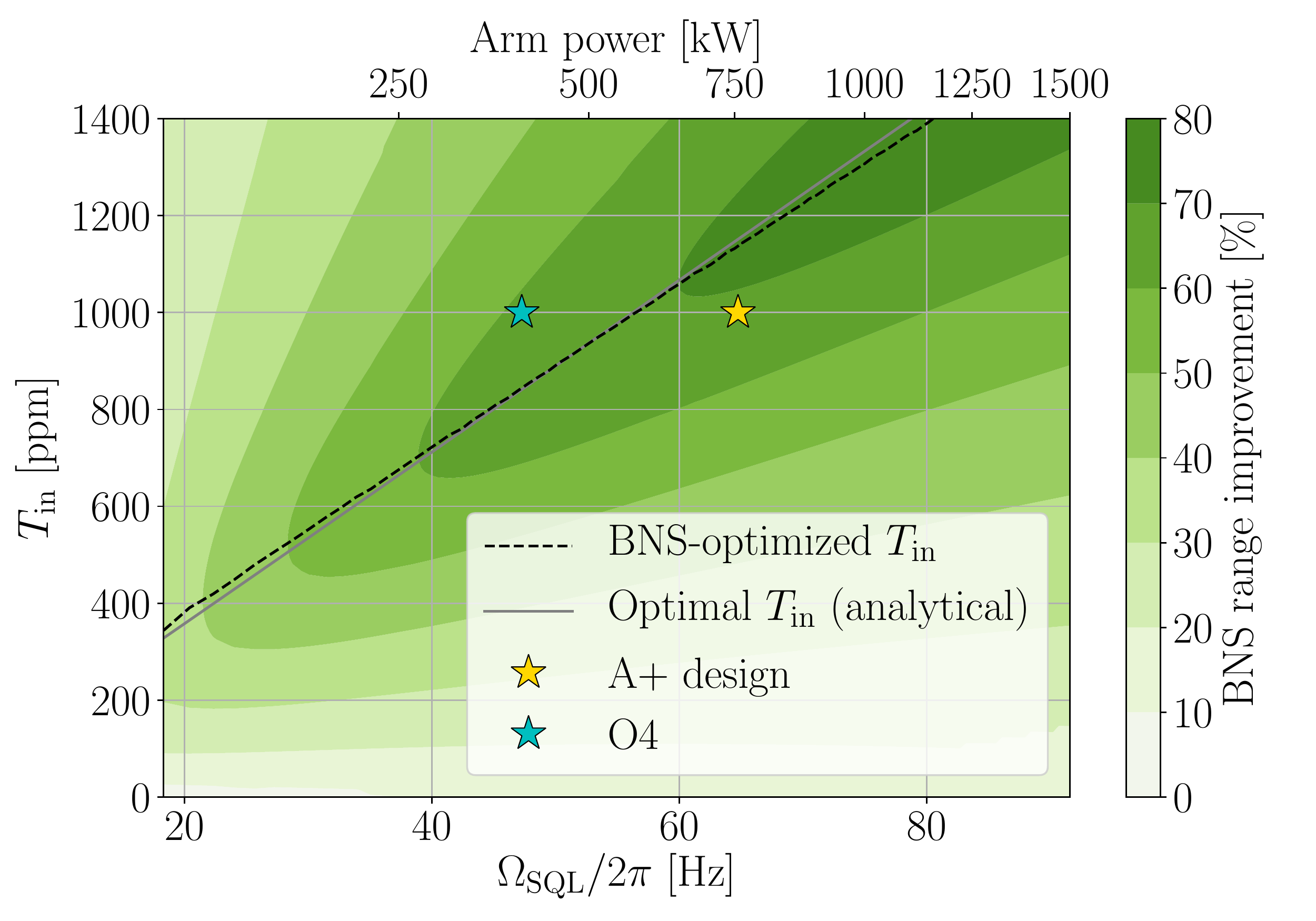}
\caption{
The relative percentage range improvement obtained with the installation of a filter cavity in A+ for various filter cavity input transmissivities and SQL frequencies. The arm power corresponding to each SQL frequency is shown in the twin axis. We assume the A+ budgeted round-trip loss $\loss = \aplusLoss$. The dashed black line indicates the optimal transmissivity at each arm power, while the gray line shows that derived from Eq.~\ref{eq:optgammadet}. The yellow star indicates the A+ design, using a filter cavity with $\Tin = \aplusT$ and a \armPower{} arm power. The cyan star marks the minimal operating power planned for the fourth observing run (O4), \SI{400}{kW}, at least twice that measured for O3. This plot shows that a \aplusT{} input coupler is close to optimal for a wide range of prospective arm powers.
}
\label{fig:trans_power}
\end{minipage}
\end{figure*}

In order to investigate the flexibility of a filter cavity with a fixed input transmissivity, we calculate the enhancement factor with changing $\Delta \omegaFC$ and $\lambda$ in Fig.~\ref{figure2}. The loss-less filter cavity with $\Delta \omegaFC/\OmegaSQL=1/\sqrt{2}$ gives the most squeezing. The black dashed line shows the optimal detuning of the cavity at each optical loss. The optimal detuning decreases with increasing optical losses. This trend is continuous toward a zero-detuning cavity: namely, the ``amplitude filter cavity'', which operates by replacing the squeezed state by unsqueezed vacuum at low frequencies~\cite{Corbitt2004,Komori2020}. The gray line shows the analytical optimal detuning in Eq.~\ref{eq:optgammadet}. 
Despite deriving our analytical model assuming ideal conditions, it also appears valid in more general scenarios of filter cavity operation.

\section{Application to the A+ filter cavity}
\psub{How we study A+}
Here we apply the above formalism to A+ as a worked example. The upgrade to Advanced LIGO will deploy a \SI{300}{m} long filter cavity with a budgeted \aplusLoss{} round-trip loss. An input optic with \aplusT{} transmissivity was chosen to optimize detector performance up to the target A+ arm power of \armPower{}. We now explore this choice of input coupler, as well as the optimal detuning of the filter cavity. We use the binary neutron star (BNS) inspiral range of the detector as our metric of performance, defined as the distance to a coalescence of two $1.4 M_\odot$ neutron stars that is detected by the interferometer with an SNR of 8, averaged over the entire sky~\cite{bnsrange}. A+ is designed to reach a BNS range of \SI{345}{\mega pc}. Similar to Eq.~\ref{eq:integration} used in Figs.~\ref{figure1} and~\ref{figure2}, this metric uses an $\Omega^{-7/3}$-weighted integration, but now casts the result in an astrophysical context. We now also include the full array of A+ classical noise curves alongside the varying quantum noise. We present this as a percentage improvement over the range of an equivalent interferometer with only frequency-independent squeezing: a \SI{100}{\percent} increase in range corresponds to a \SI{6}{\decibel} enhancement of sensitivity.

\begin{figure*}
\centering
\begin{minipage}[t]{0.48\hsize}
\vspace{0pt}
\centering
\includegraphics[width=\hsize]{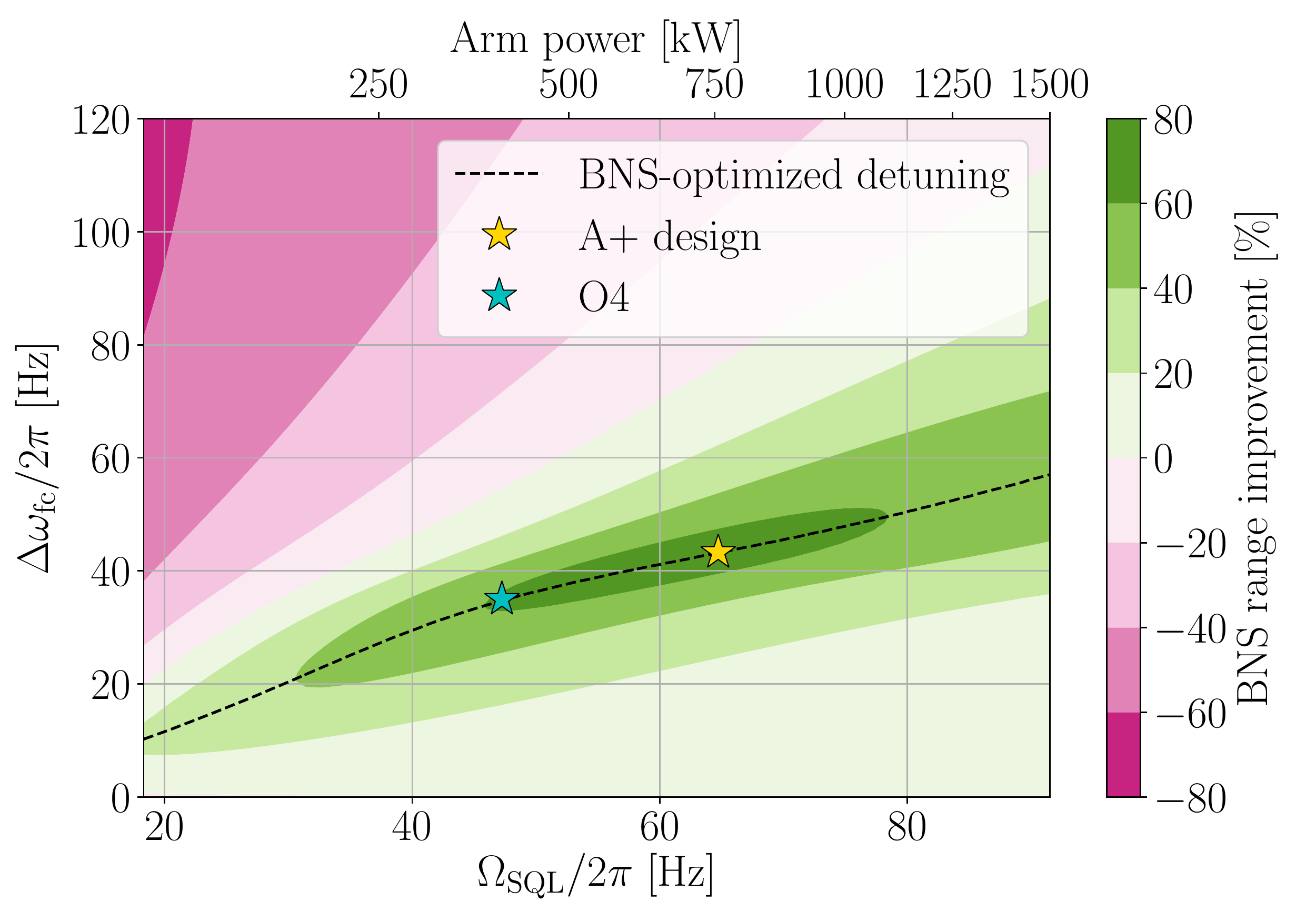}
\caption{
Relative percentage binary neutron star (BNS) inspiral range improvement gained by installing a filter cavity with A+ parameters ($\Tin = \aplusT{}$, $\loss=\aplusLoss{}$) at various interferometer SQL frequencies. The arm power corresponding to each SQL frequency is shown in the twin axis. The vertical axis explores possible filter cavity detunings, with the black dashed line highlighting the optimal such value for each $\OmegaSQL$. The yellow star marks the designed A+ arm power. We see that small adjustments in operating point can be used to mostly compensate for deviations from the designed arm power from \SI{50}{\hertz} to \SI{70}{\hertz} SQL frequencies.
}   
\label{fig:SQL_fdetune}
\end{minipage}\hfill
\begin{minipage}[t]{0.48\hsize}
\vspace{0pt}
\centering
\includegraphics[width=\hsize]{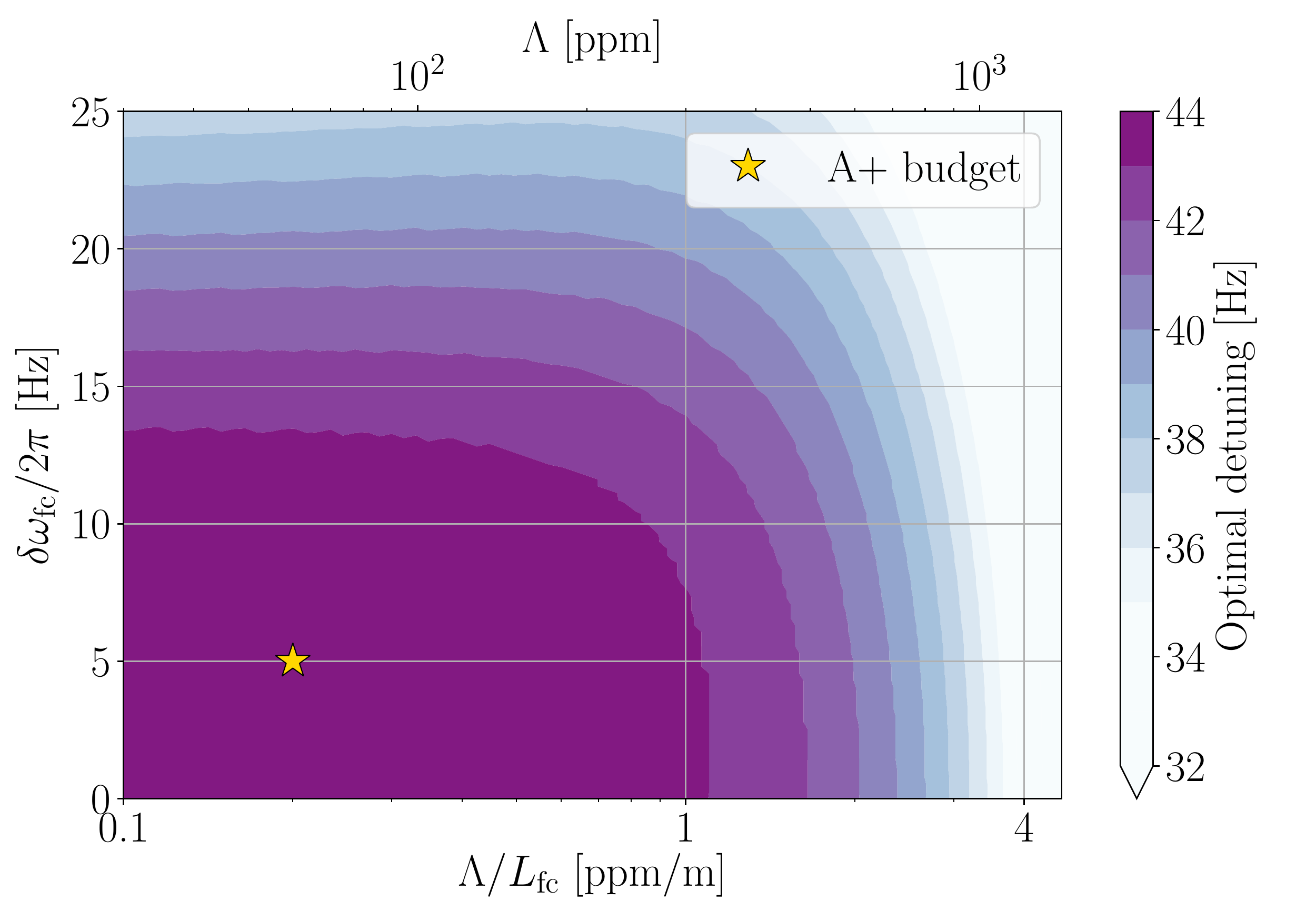}
\caption{
The required detuning to accommodate increasing losses or detuning fluctuations. The yellow star marks the budget for the A+ filter cavity design. For any realistic round-trip loss or detuning noise, only small changes in $\Delta \omega_\text{fc}$ are required.
}   
\label{fig:aplus_degradation}
\end{minipage}
\end{figure*}

\psub{Choosing filter cavity transmissivity}
We first explore the selection of a filter cavity input transmissivity.
Fig.~\ref{fig:trans_power} shows the optimal choice for varying interferometer arm powers and, correspondingly, SQL frequencies.
The range improvement for a given arm power and input coupler is calculated by optimizing the filter cavity detuning and squeezing level.
This plot is particularly relevant for interferometers undergoing iterations of upgrades; the usage of highly-transmissive input couplers to target high-power operation penalizes operation at lower powers.
For Advanced LIGO progressing into A+, we see that a choice of \aplusT{} is within \SI{5}{\percent} of optimal BNS range for arm powers in the range \SIrange{400}{800}{kW}.
We also consider this choice with a signal recycling cavity adjusted for higher power in Appendix~\ref{app:srm}.
Upon realizing the A+ design, the filter cavity input optic can be optimized for long-term observation at the final interferometer configuration. 
Finally, we note that the analytical solution from Eq.~\ref{eq:optgammadet} is valid in this regime.

\psub{Dealing with different powers}
Suppose we use Fig.~\ref{fig:trans_power} to design our filter cavity and arrive at the A+ design input transmissivity of \aplusT{} with a budgeted loss of \aplusLoss{} and a \aplusLength{} filter cavity (\SI{0.2}{ppm\per\metre}). We now consider sensitivities achieved from the application of this fixed filter cavity to interferometers with varying SQL frequencies. Fig.~\ref{fig:SQL_fdetune} shows how the filter cavity detuning should be adjusted to compensate for interferometer powers for which the cavity was not designed. The detuning must change to approximately offset the varying interferometer rotation frequency. As expected, the greatest improvement in performance---nearly doubling the BNS inspiral range---is reserved for the designed interferometer power.

\psub{Filter cavity detuning noise}
The second derivative of range with respect to detuning gives a scale for the loss of sensitivity due to detuning noise.
We can use the narrow region of optimum performance in Fig.~\ref{fig:SQL_fdetune}, centered on the dashed line, to infer a detuning noise requirement for the filter cavity.
For instance, if we demand that the standard deviation of the detuning remains within \SI{1}{\percent} of the maximum range, we derive an upper limit for the detuning noise RMS of \SI{1.2}{\hertz} (\SI{1.3}{\hertz}) for A+ (O4), or equivalently an effective length noise RMS of \SI{1.3}{\pico\metre} (\SI{1.4}{\pico\metre}).
The A+ filter cavity design chooses a detuning noise constraint that limits the injected anti-squeezing noise to be no greater than the squeezed shot noise itself, leading to a slightly more restrictive length noise upper bound of \SI{0.8}{\pico\metre}~\cite{AplusDesign}.
Ref.~\cite{Kwee2014} discusses detuning noise requirements based on the resulting frequency-dependent phase noise in more detail.

\psub{Dealing with different loss/detuning noise}
We additionally explore the required detuning shifts to compensate for more severe squeezing degradation. In particular, Fig.~\ref{fig:aplus_degradation} shows the required change in filter cavity detuning given varying round-trip loss and detuning fluctuations. In general, we find that worse filter cavities require operation closer to resonance. This is consistent with the trend shown in Fig.~\ref{figure2} toward amplitude filter cavity operation: the greater the mismatch between the filter cavity and interferometer rotations, the more one benefits from running the cavity as a high-pass filter. Further, we note that the A+ design filter cavity is highly tolerant of a range of degradations, requiring a detuning shift of only a few Hz for losses of up to a few hundred ppm and $\sim\SI{15}{\hertz}$ of detuning noise. Compare these to the measured values of \mitLoss{} and \mitDetuningNoise{} measured using a \cavityLength{} filter cavity with a comparable bandwidth in Ref.~\cite{McCuller2020}.

\section{Conclusions}

Present-era gravitational-wave detectors are just now beginning to become radiation pressure noise limited. Frequency-dependent squeezing upgrades will imminently allow us to circumvent this limit and achieve a broadband reduction in quantum noise. The design and operation of a filter cavity requires a careful treatment of noise terms and degradation mechanisms. In this paper, we report on a more general and simplified expression of the sensitivity improvement possible with frequency-dependent squeezing.

We also explore the parameter space of filter cavities to determine an optimal operating point given some budgeted loss, maximizing on an inspiral-weighted integrated spectrum. With the full A+ design in mind, we similarly study detuning changes to compensate for mismatched rotation frequencies, as well as excess loss and length noise. This study aims to be valuable when operating a filter cavity throughout a phased upgrade from current to future detectors with more arm power and stronger squeezing.

\section*{Acknowledgements}
LIGO was constructed by the California Institute
 of Technology and Massachusetts Institute of Technology with funding from the
 National Science Foundation, and operates under Cooperative Agreement No. PHY-1764464. Advanced LIGO was built under Grant No. PHY-0823459. KK is supported by JSPS Overseas Research Fellowship.
 
 We would like to thank Haixing Miao and Eugene Knyazev for useful comments.

This paper has LIGO Document Number LIGO-P2000240.

\clearpage\newpage

\begin{figure*}[!t]
\centering
\begin{minipage}[t]{0.48\hsize}
\vspace{0pt}
\centering
\includegraphics[width=\hsize]{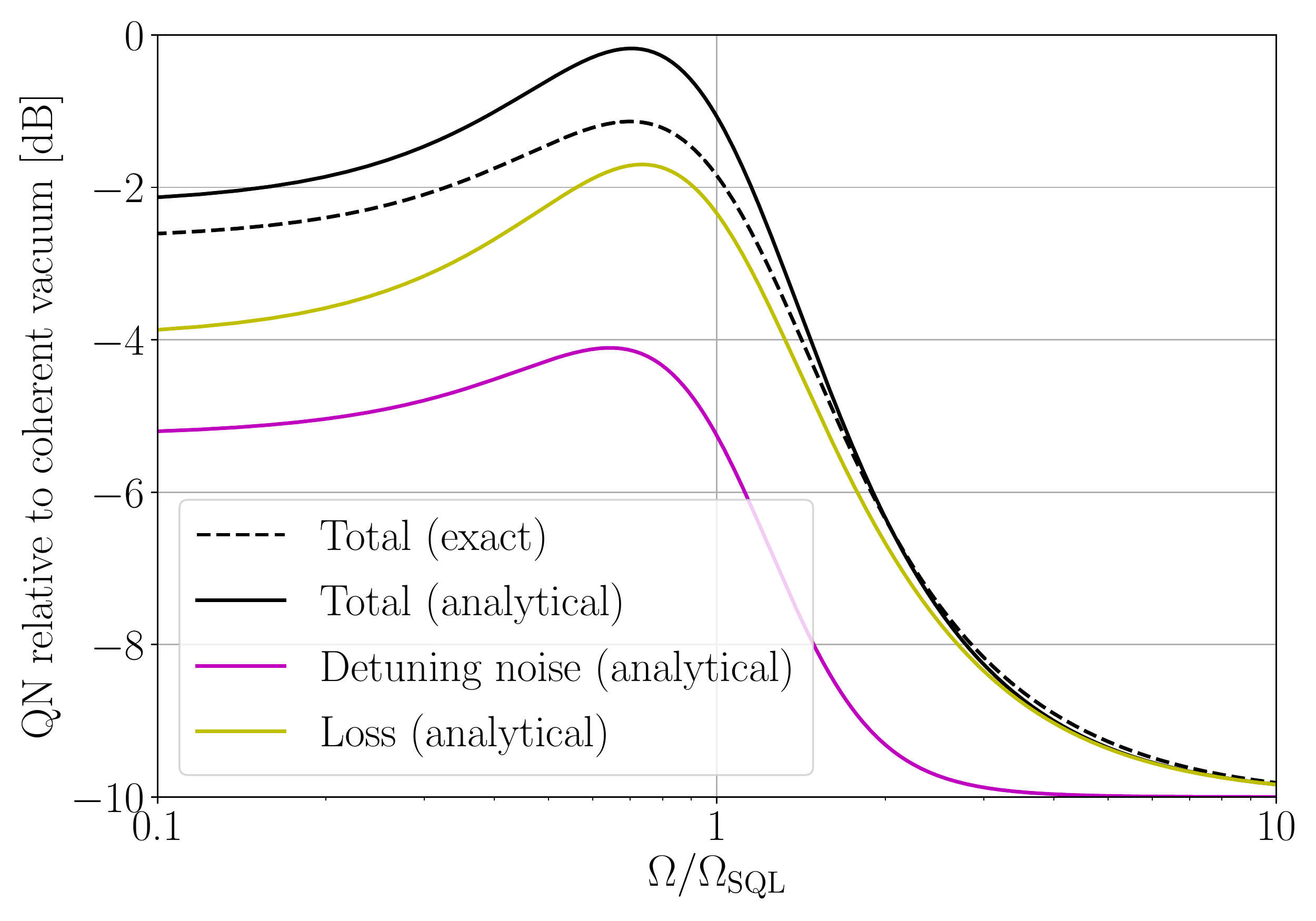}
\caption{
Quantum noise relative to coherent vacuum calculated to only each leading order term alongside the exact numerical simulations. Parameters used are the same as those in the middle panel of Fig.~\ref{figure1}: $\e^{-2\sigma}=0.1$, $\ell=0.15$, and $\xi=0.1$. The analytical equation shows better agreement with the exact result at frequencies higher than $\OmegaSQL$, while it overestimates the noise at lower frequencies.
}
\label{figureapp}
\end{minipage}\hfill
\begin{minipage}[t]{0.48\hsize}
\vspace{0pt}
\centering
\includegraphics[width=\hsize]{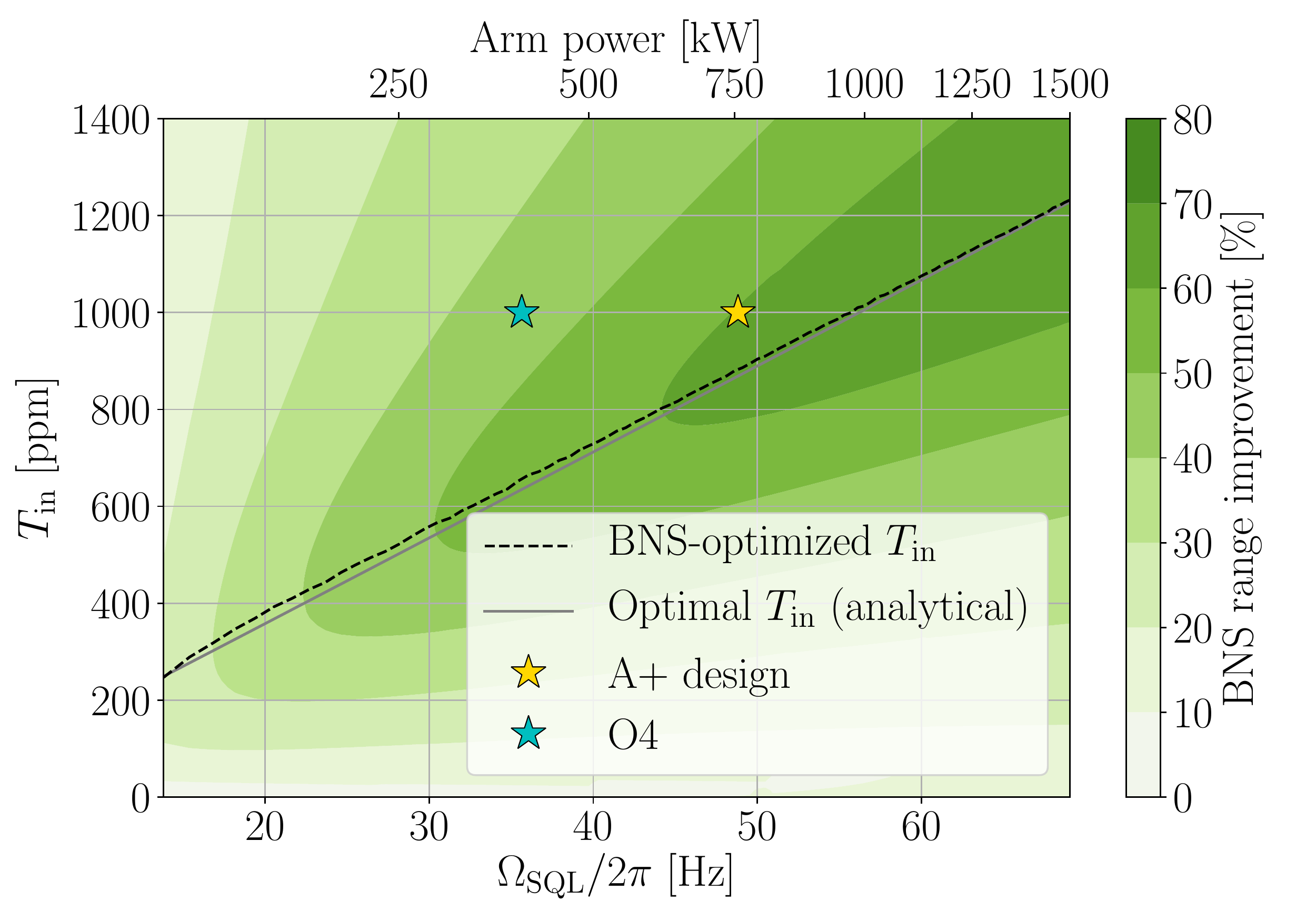}
\caption{
The range improvement as a function of arm power and filter cavity input transmissivity, now using an SRM transmissivity of \srmDesign{}.
We see that \aplusT{} is again a good choice for this configuration, giving near-optimal range improvement for powers at and extending beyond the A+ design goal.
}
\label{fig:srm}
\end{minipage}
\end{figure*}
\appendix
\section{Analytical formula}
\label{app:formula}
Here we derive new analytical formulae that now incorporate both optical loss and detuning fluctuations. We follow the approach of Ref.~\cite{Kwee2014} in averaging over these fluctuations. The phase matching condition in Eq.~\ref{eq:optgammadet} is used in order to determine the optimal detuning and transmissivity. Assuming $\ell, \xi \ll 1$ and $\K \simeq (\OmegaSQL / \Omega)^2$, the overall change of quantum noise relative to that without injecting squeezed vacuum, given by $s=1 + \K^2$, can be represented as
\begin{align}
\label{eq:lossphasedegradation}
\frac{\Ntot}{s} &= \e^{-2\sigma} + \frac{2\sqrt{2}\K (1 + \K)}{1 + \K^2} \left( 1 - \e^{-2\sigma} \right) \ell \nonumber \\
&+ \frac{2\K^2(1 + \K)^2}{(1 + \K^2)^2} \left( \e^{2\sigma} - \e^{-2\sigma} \right) \xi^2 \nonumber \\
&+ \frac{4\K^2}{(1 + \K^2)^2} \left[ \K (\e^{2\sigma} - 1) \right. \nonumber \\
&\left. \hspace{2cm} + (\K^2 + 3\K + 1)(\e^{-2\sigma}-1) \right] \ell^2 \nonumber \\
&+ \mathcal{O}(\ell \xi^2) + \mathcal{O}(\ell^3) + \mathcal{O}(\xi^4).
\end{align}
Eq.~\ref{eq:lossphasedegradation} goes to unity when the injected light is vacuum ($\sigma =0$), regardless of $\ell$ and $\xi$. The leading term of each coefficient in Eq.~\ref{eq:sqcoeff1}-\ref{eq:sqcoeff3} on $\ell$ and $\xi$ is given by
\begin{gather}
\Csq = 1, \\
\Canti = \frac{2\K^2 (1 + \K)^2}{(1 + \K^2)^2} \xi^2, \\
\Closs = \frac{2\sqrt{2}\K (1 + \K)}{1 + \K^2} \ell.
\end{gather}
While $\Canti = 0$ for an ideal cavity with optimal phase matching, we see here that the detuning noise mixes in some anti-squeezing into the quantum noise spectrum.
Both $\Canti$ and $\Closs$ are maximal at $\K_\text{worst} = \sqrt{2} + 1$, which indicates that the degradation of squeezing by the optical loss and detuning fluctuation is largest at $\Omega = 1 / \sqrt{\sqrt{2} + 1} \; \OmegaSQL \simeq 0.64 \; \OmegaSQL$.
In Fig.~\ref{figureapp}, quantum noise relative to coherent vacuum is plotted to compare Eq.~\ref{eq:lossphasedegradation} with the exact numerical calculations. In the analytical equation, we consider the terms up to $\ell^2$ and neglect higher-order ones. It approximately reconstructs the exact solution at frequencies higher than the SQL frequency. At lower frequencies, the approximated analytical calculation gives an over-estimated result, which implies that higher-order terms suppress the degradation more.

\section{Choice of signal recycling mirror}
\label{app:srm}
The existing signal recycling mirror (SRM) in Advanced LIGO was chosen to optimize for a low power operation.
Moving from \OFarmPower{} to \armPower{}, the detector becomes more limited by quantum noise than thermal noise in the $\sim$\SI{100}{\hertz} region.
Insofar as it is quantum noise limited, the range hits a maximum as radiation pressure noise and shot noise trade off due to the $\OmegaSQL$ nearing merger frequencies.

Furthermore, while we can reduce the quantum noise with frequency-dependent squeezing, quantum radiation pressure noise also acts to enhance any optical scattering noises above the squeezed vacuum, scaling with $\OmegaSQL$.
This effect motivates lowering $\OmegaSQL$ to diminish the impact of technical noises.
Such a change amounts to modifying the interferometer bandwidth,
in turn changing $\OmegaSQL$ as
\begin{gather}
\label{eq:OmegaSQL}
    \OmegaSQL = 4 \sqrt{\frac{\Parm \omegaCarrier}{ c m \Larm \gammaIfo}},
\end{gather}
where $\Parm$ is the intra-cavity arm power, $\omegaCarrier$ is the carrier frequency, $m$ is the test mass and $\Larm$ is the arm length.
These factors additionally manifest in the overall differential displacement due to quantum noise as~\cite{Yu2020}
\begin{equation}
    \Delta x^2(\Omega) = N_\mathrm{tot} \frac{\gammaIfo^2 + \Omega^2}{\gammaIfo} \frac{\hbar c \Larm}{4\omega_0 \Parm}.
\end{equation}

At design power, the SRM transmission is planned to decrease from \srmLowPower{} to \srmDesign{}~\cite{Aasi2015}, bringing the interferometer bandwidth from \ifoBandwidth{} to \ifoBandwidthSRMchange{}.
As a result, $\OmegaSQL$ is shifted from \sqlFreq{} down to \sqlFreqSRMchange{}.
In Fig.~\ref{fig:srm}, we show that the choice of $\Tin = \aplusT$ similarly achieves great range improvement for the A+ design in this configuration, as well as even higher arm powers.

\bibliography{paper}
\bibliographystyle{apsrev.bst}
\end{document}